\documentclass[draft]{agujournal2019}
\usepackage{url} 
\usepackage{lineno}
\usepackage[inline]{trackchanges} 
\usepackage{soul}
\draftfalse

\journalname{Radio Science}

\begin{document}

%
%


\title{Seasonal and diurnal dynamics of radio noise for 8-20MHz poleward-oriented mid-latitude radars}

%
%




\authors{O.I.Berngardt\affil{1}, J.-P.St.Maurice\affil{2}, J.M.Ruohoniemi\affil{3}, A.Marchaudon\affil{4}}

\affiliation{1}{Institute of Solar-Terrestrial Physics, SB RAS, Irkutsk, Russia}
\affiliation{2}{Department of Physics and Engineering Physics, University of Saskatchewan, Saskatoon, Canada}
\affiliation{3}{Bradley Department of Electrical and Computer Engineering, Virginia Polytechnic Institute and State University, Blacksburg, VA, USA}
\affiliation{4}{Research Institute in Astrophysics and Planetology (IRAP), Toulouse University, CNRS, CNES, Toulouse, France}





\correspondingauthor{Oleg I.Berngardt}{berng@iszf.irk.ru}




\begin{keypoints}
\item Physics-based seasonal-daily noise level model  built for pole-oriented HF radars
\item Good agreement between  modeled noise level and experimental observations
\item Model allows estimating vertical absorption from noise measurements
\end{keypoints}

%
%

%
%


\begin{abstract}

Based  on ray tracing in a smooth ionosphere described by the IRI-2016 model 
we have infered the seasonal-diurnal dynamics of radio noise observed by four 
mid-latitude HF radars. In the calculations, noise is assumed to propagate 
from the radar dead zone boundary.
Noise absorption along the ray path is simulated from  the IRI-2016 electron density, and from
the molecular nitrogen density
and electron temperatures obtained from the NRLMSISE-00 model.
Model results are compared with experimental
radar data, and good agreement between the two is demonstrated.
The model  makes it possible to estimate the amount of absorption
in D- and E- layers under average undisturbed conditions.  This is
important for the retrieval of long term variations in the electron density
in the lower ionosphere.  The model also makes it feasible  to interpret vertical absorption in
experimental data, thereby significantly expanding the capability of HF  radars to monitor  the lower ionosphere.

\end{abstract}


%
%

%


%
%
%
%

\section{Introduction}

Noise level studies have recently become an integral part of
theoretical and applied research on lower ionosphere dynamics
during solar flares and coronal mass ejections through the use of SuperDARN and similar high-frequency
(HF) radars \cite{Berngardt_2018,Bland_2018,Berngardt_2019,Bland_2019,Berngardt_2020}.
 HF noise has traditionally been considered to come from a mixture of several
components - anthropogenic, cosmic, and atmospheric.

For  radars with an equatorward field of view, it has been  assumed that the main contribution to observed
HF noise originates from tropical lightning discharges and that the signals from these discharges
propagate over large distances through the ionospheric waveguide  \cite{Pederick_2014}.
Modeling based on this notion has been validated reasonably well, with  noise levels 
predictions close to experimental observations.

\begin{figure}
\includegraphics[scale=1.2]{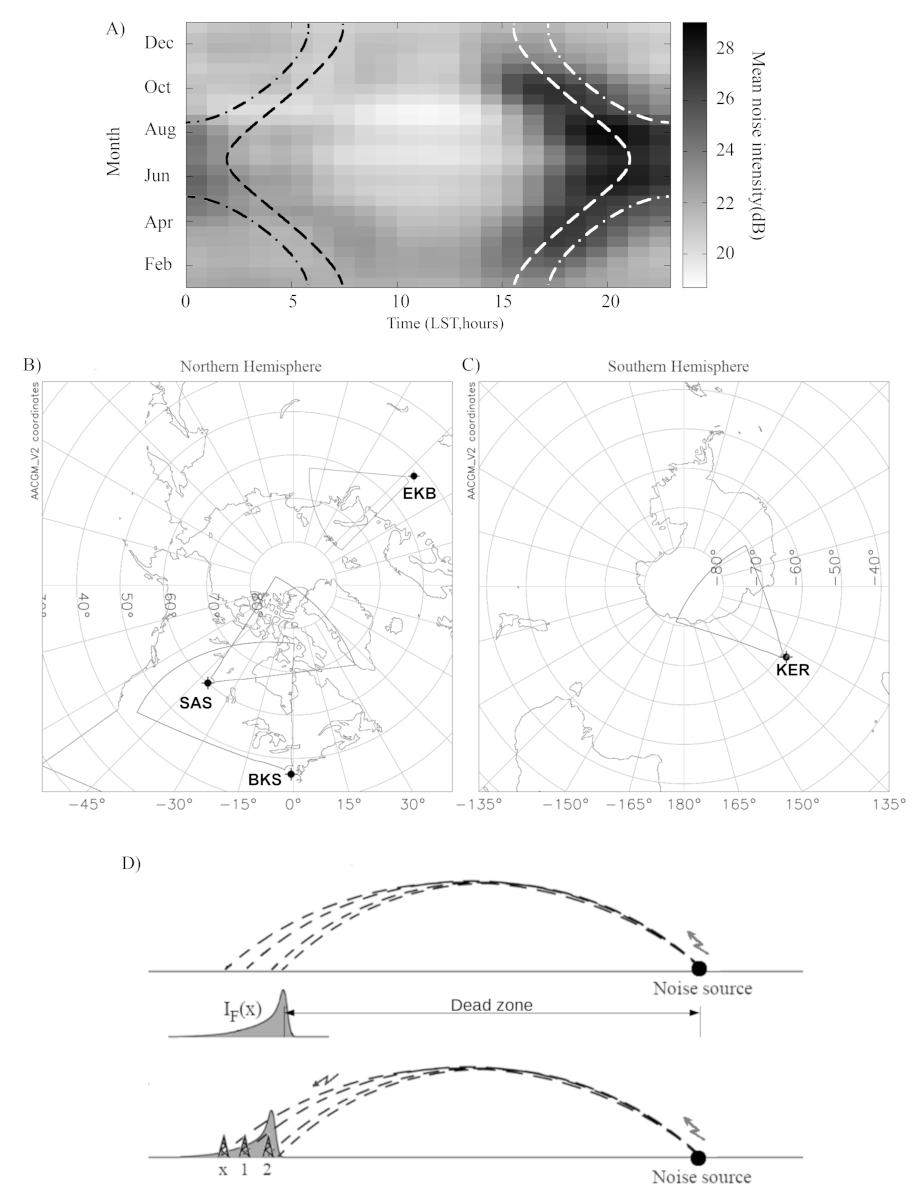}

\caption{A) Seasonal-diurnal pattern of average noise intensity at the EKB radar  \cite{Berngardt_2020}.  
The lines correspond to the position of solar terminator at two different heights 
(at ground level - dashed line and at 300km height - dot-dashed line);
B-C) Fields of view,  in AACGMv2 \cite{Shepherd_2014} geomagnetic coordinates, of the four mid-latitude
radars involved in the current study, for the Northern and Southern hemispheres;
D) Noise propagation and collecting scheme for the proposed model, illustrating the focusing 
mechanism for enhancing the intensity, $I_F(x)$, around the boundary of the dead zone.
}

\label{fig:fig1}
\end{figure}

In  studies by \cite{Ponomarenko_2016} and \cite{Berngardt_2020}, seasonal-diurnal variations 
of the noise level were measured by the mid-latitude
SAS and EKB radars at a fixed frequency. 
In this paper we refer to radars, including SAS, as mid-latitude radars 
for having geographic latitudes lower than 60 degrees.
The fields of view of both radars point 
toward the polar regions and the shapes of the seasonal-diurnal dependence as a
function of local solar time are similar.  As illustrated in  (Fig.\ref{fig:fig1}A), 
they take the form of an oval, the contour of which is determined by the position of 
the solar terminator.  This stated, there is no obvious reason why the HF noise observed in 
mid-latitude poleward pointing radars like SAS and EKB should be closely associated with 
lightning.  In addition, the ray paths emerging from equatorial sources are 
so long that the noise should be absorbed almost completely.

 As a result of the above considerations, \cite{Berngardt_2019}  suggested that 
the observed noise pattern
was not so much related to the source of noise in poleward pointing mid-latitude 
HF radars as it was related to ionospheric conditions.  The authors proposed
that in most cases the  sources of noise should  be considered to be anthropogenic 
and that the noise pattern is dictated by
the border of the so-called 'dead zone' often observed though  ionospheric ground 
echoes \cite{Samson_1990}. The dead zone is an area around the radar that is free of 
hop-propagating rays in terms the geometric optics approach 
that can complete a path between a ground point and the radar, hence, 
points within the dead zone are not sources of noise for the radar (
in terms of wave optics approach, 
used in this paper, some weak signals can be detected even in the dead zone ).
Dead zones mark  the boundary of a noise region 
where ray paths are focused by ionospheric effects.   Specifically, this means 
that both the trajectory of
propagation of noise signal and the intensity of the noise focusing at the
border of the dead zone depend on the height distribution
of the electron  density along the ray path, which has clear and strong 
diurnal and seasonal variations.
Based on that notion, a  model of the noise pattern for poleward pointing 
radars at mid-latitudes would have to properly account for the seasonal 
and daily variations of the noise level through the proper use of ionospheric 
and atmospheric models, given that the noise and its pattern both depend on 
the atmospheric  and ionospheric densities through absorption and ray path 
focusing/de-focusing, respectively.

Mathematical empirical approaches have been introduced to reproduce  noise patterns of the 
kind shown in Fig.\ref{fig:fig1}A, irrespective of the physical origin of the patterns.  
One such approach uses the mean daily dynamics of noise level
over the previous 30 days \cite{Bland_2018}.  Another consists of using special 
auto-regressive models \cite{Berngardt_2020}.  While these models have their usefulness,  
no physical model has apparently been built to this day to reproduce  the seasonal-daily 
dependence of the noise pattern.  One clear advantage of building such a physical model 
would be the ability to predict a 'quiet-day' curve, namely, a curve of the expected noise level for
ordinary ionospheric conditions. Reproducing an observed quiet-day curve would make it
possible to estimate the vertical absorption level, allowing for the extraction of
the long-period variations  associated with the D- and E-layers of the ionosphere.
In turn, building a physical model of noise level would make it feasible
to effectively use SuperDARN and similar radars for estimating the electron
density in the lower part of the ionosphere.

With the above in mind, we have developed a physics-based numerical model of the noise at HF 
frequencies.  We have applied the results of the model to the seasonal-diurnal dynamics of
noise level at four mid-latitude radars, namely,  the Canadian SAS SuperDARN radar  (52.2N, 106.5W), 
the US BKS SuperDARN radar (37.1N, 78.0W), the French KER SuperDARN radar (Southern French Territory, 49.2S, 70.1E)
and the Russian EKB ISTP SB RAS radar (56.5 N, 58.5E). Each of these radars has a field of view with an
antenna beam pattern that points nearly poleward (Fig.\ref{fig:fig1}B-C), meaning that
the influence of tropical thunderstorm centers on the noise production should  be negligible.  
We describe our model in Section 2.  In Section 3 we present its results and compare  with 
observations while discussing the implications for the plasma densities in the lower ionosphere.  
Our conclusions are presented in Section 4.

\section{Description of the physics-based noise  model}

Our model is based on a mechanism proposed by
\cite{Berngardt_2018,Berngardt_2019}. It assumes that the noise level in polar-oriented radars 
operating in the 8-20 MHz
range is determined by a focusing of noise signals from various
anthropogenic sources, and that  the main sources contributing to the noise 
are located around the boundary of the so-called 'dead zone'.   Near the dead zone,
the amplitude of propagating signals is  amplified significantly
due to spatial focusing of radiowaves by the ionosphere \cite{Tinin_1991}.

Within the framework of this model, the following factors
need to be considered:
\begin{enumerate}
\item the spatial distribution of the amplitudes of the sources of  noise and
the noise emission direction pattern for each source;
\item the ray path of the radio signals in an irregular ionosphere;
\item the intensity produced by the focusing of radio signals on the boundary of the dead zone;
\item the absorption of the signal over  the ray path as it passes through the D- and E-regions;
\item the antenna beam pattern of the radar receiver.
\end{enumerate}
In  general, each of the above characteristics has its own temporal and frequency
dependencies. This stated, in the following calculations, we assumed that noise sources approximately have
an isotropic directional pattern, equal amplitudes, and are uniformly
distributed over the Earth's surface around the radar.
In the presence of intense localized
anthropogenic sources (radio stations, industrial sources, railways)
or for radars facing  thunderstorm activity centers, this assumption has to break down. 
However, for radars located at a significant
distance from intense anthropogenic or thunderstorm activity centers, this
assumption should be acceptable.

To be specific, according to the interpretation suggested by \cite{Berngardt_2018,Berngardt_2019,Berngardt_2020}
noise  propagates from a noise source and
focuses just beyond the boundary of the source's dead zone,   as illustrated in Fig.\ref{fig:fig1}D. 
Within that framework the noise received by a radar
is a superposition (integral) of noise source contributions with weights defined by
dead zone multiplier $I_{F}$ calculated for the radar position. The integration is made over
the azimuth $\theta$, elevation $\alpha$, and range $s$. It takes into account the antenna 
beam pattern, signal attenuation
with distance and signal attenuation due to absorption in the lower ionosphere.
This means that, to a first approximation, the noise intensity received by the radar, 
${I_{rcv}}$,  can be obtained from the expression
\begin{equation}
{I_{rcv}} \approx \int_{\Omega,S} I_{src}\left [x(\alpha,\theta)\right]I_{F}\left [x(\alpha,\theta,s)-x_0(\alpha, \theta )\right ]A_{t}(\alpha,\theta)G(\alpha,\theta)s^{-2} \cos(\alpha) d\alpha d\theta ds
\label{eq:initial_equation_1}
\end{equation}
where $I_{src}$ is the  noise intensity distribution of the noise source  at the ground, 
$I_F$ is the dead zone focusing multiplier,   $s$ is the distance from the source to 
the receiver, and $s^{-2}$ is the attenuation in intensity due to distance.  
Also, $G(\alpha,\theta)$ is the shape of the antenna beam pattern, and $A_t(\alpha,\theta)$ 
is the absorption over the trajectory of a ray. In these expressions $x$ is a phase path 
length along a ray that starts at specific values of $\theta$ and $\alpha$, whereas $x_0$ 
is the phase path length at the focusing point, namely, at the dead zone boundary.  
It should be understood that $x_0$ does not depend on $s$ but that $x$ does depend 
on $s$ because the integration is over the distances of the various  noise sources 
to the receiver. In other words, the   $\alpha$ and $\theta$ integrations  are  over 
the receiver angles, while the  integration over $s$ is actually  over the distance 
of the noise sources to the radar.

As a further  approximation, we assume that the variations of the various parameters are such that the noise intensity
$I_{rcv}$ can   be  described accurately enough with the simpler expression
\begin{equation}
{I_{rcv}} \approx I_{0}\overline{I_{F}}A_{t}G_{A}\overline{A_R}
\label{eq:initial_equation}
\end{equation}
where $I_{0}$ is the intensity of ground noise sources and replaces the $I_{src}$ if 
the latter can be assumed to be isotropic (later in our simulations 
$I_0$ is used as a normalization constant, calculated over one year). 
In addition, $\overline{I_{F}}$
is the integral over all distances to the source of the focusing intensity at 
the boundary  of the dead zone, whereas $A_{t}$ becomes the absorption
of the radio signal power over the whole ray path, and $G_{A}$ is the beam pattern 
of the receiving antenna
 integrated over $\theta$, i.e., over all azimuths. Also, $\overline{A_R}$ is the 
integrated signal attenuation with respect to distance.
Consistent with this definition, the $\overline{a}$ symbol describes  an averaging of quantity $a$ over the range.
So to build Eq.\ref{eq:initial_equation} we replace average value of the product of 
the various non-negative parameters in Eq.\ref{eq:initial_equation_1} by 
the product of average values of the parameters with a single ray path
from the border of the radar dead zone.

Finally, taking into account that power attenuates with propagation trajectory length $s$ as $s^{-2}$,
we can estimate:

\begin{equation}
\overline{A_R} \approx \int_{S}^{S_{max}} s^{-2}ds \approx S^{-1}
\end{equation}
This result is valid as long as $S << S_{max}$ (maximal radar range).

At a fixed frequency, we can assume $I_{0}$ to be constant. This holds if we can 
assume to first order that the various noise sources are not only isotropic, but also homogeneous,
incoherent, and equal in intensity. This is a rather rough assumption,
but, as shown below, it proves to be  sufficient for an adequate modeling of the noise pattern.

\subsection{The propagation trajectory}

Propagation trajectories were obtained with the use of wave optics based on 
a calculation made at fixed pre-determined starting values of the elevation 
angles   within 0-90 degrees with
0.09 degree steps.  For a given azimuthal direction or radar beam we could
describe the system in terms of spherical coordinates for the radial distance 
$\rho$ and the elevation angle $\phi$.  We restricted the calculations to 
the high-frequency
limit for the index of refraction, $n$, and  did not include absorption 
in the calculations, i.e., we used the ordinary wave  expression given by

\begin{equation}\label{eq:jp-index}
    n^2= 1- f_e^2/f_0^2
\end{equation}
where   $f_e$ and $f_0$ are  the plasma frequency and radar frequency, respectively. 
For this situation  the differential equations leading to phase and group delays as 
a function of starting elevation angles can be described by the following set of 
equations \cite{Kravtsov_1983}:

\begin{equation}
\left\{ \begin{array}{l}
\frac{\partial R_{\rho}}{\partial s}=P_{\rho}\\
\frac{\partial R_{\phi}}{\partial s}=\frac{1}{R_{\rho}}P_{\phi}\\
\frac{\partial P_{\rho}}{\partial s}=\frac{f_{e}}{f_{0}^{2}}\frac{\partial f_{e}}{\partial R_{\rho}}+\frac{1}{R_{\rho}}P_{\phi}^{2}\\
\frac{\partial P_{\phi}}{\partial s}=\frac{1}{R_{\rho}}(\frac{f_{e}}{f_{0}^{2}}\frac{\partial f_{e}}{\partial\phi}-P_{\rho}P_{\phi}) \\
\psi=\psi_{0}+\int\limits _{S_{0}}^{S}P^{2}ds
\end{array}\right.\label{eq:num_model_eq-1}
\end{equation}

where $\overrightarrow{R}$ is the radius vector describing the position 
of the wave impulse at a particular point in the trajectory  and where  
$\overrightarrow{P}$ is the wave impulse, while $\psi$ is the
phase path (or eikonal). Integration was carried  over the trajectory 
variable $s$  with starting and end points at $S_0$ and $S$, and with $\psi_{0}$
being associated with the initial value of the eikonal.
In such calculations, the regions where phase delay
goes through an extremum (minimum) correspond to the regions
of  focusing, that is to say, they describe the borders of the dead zone for
different propagation modes.

In the numerical calculations giving rise to our model,  we assumed that 
the radio wave propagated in
a plane that was associated with a given azimuth over a globe arc within
the framework of the spherical Earth model.  This stated, our ionospheric model remained
inhomogeneous in the propagation plane, namely, over the distance to the source and the elevation angle.
In other words the ionospheric model remained a function of latitude and altitude.  
Specifically, we used the IRI-2016 \cite{Bilitza_2017} ionospheric model for this purpose.
In order to work with  a smooth and continuous ionosphere along the trajectory, 
it was necessary to assume that the signal propagation plane did not depend on
the starting elevation angle when seeking numerical solutions to the set of 
differential equations (\ref{eq:num_model_eq-1}). While correct  for 
the middle(central) beams
of the radar antenna pattern, this was not strictly true for the extreme
ones, since the antenna beam pattern is actually described by  a conical shape \cite{Shepherd_2017}. But, as  will
be shown later, this rough approximation is sufficient for carrying 
the model calculations forward with reasonable accuracy.

For the specific ray path calculations we used  a grid with
 step sizes of 222 km (equivalent to $2^o$ in latitude at Earth's surface) 
in distance and 3 km in height in the propagation plane.
 For calculations
between grid points, an interpolation was performed using two-dimensional
local quadratic B-splines to obtain continuous values and smooth
spatial derivatives.
When later adding absorption effects we used  the
 the NRLMSISE-00 model for the neutral atmosphere, from which we extracted the
 molecular nitrogen density and the electron temperature interpolated between grid points in the same way.

\subsection{Modeling the focusing near  the dead zone boundary}

\subsubsection{Determining $\overline{I_{F}}$}

In a two-dimensional situation such as the one used here, the
focusing of the radio waves emitted from a point source can 
be described analytically near the dead zone, if the ionospheric 
peak can be approximated by a parabola.
In that case, if we neglect insignificant multipliers,
the signal amplitude $A_{F}(x)$ over the phase path length $x$
has a well-known dependence given by \cite{Tinin_1991}:

 \begin{equation}
A_{F}(x)\sim\frac{1}{\Lambda}Ai(\frac{2^{2/3}}{\Lambda}[\overline{x}-x])
\label{eq:IF_partial}
\end{equation}

where $Ai()$ is the Airy function describing the shape of the signal
focusing versus phase path length, and where

\begin{equation}
\Lambda=\left(\frac{1}{2k^{2}}\frac{\partial^{2}\overline{x}}{\partial \alpha^{2}}\right)^{1/3}
\label{eq:lamda}
\end{equation}

is a quantity inversely proportional to the size of the focusing area
associated with the second derivative of the phase path length over the elevation
angle $\alpha$. Note that  Equation (\ref{eq:IF_partial}) does not take into account
the  decay of the wave field with distance.

By integrating (\ref{eq:IF_partial}) over the range, and taking into account that noise
sources are independent, it therefore follows
that the range-integrated power $\overline{I_{F}}$, corresponding
to the total power of the noise superposed from various noise sources homogeneously placed at  different
ranges is given, to a first approximation, by:

\begin{equation}
\overline{I_{F}}\sim \frac{1}{\Lambda} \int Ai^2(\overline{\eta}-\eta)d\eta
\label{eq:IF_Integral}
\end{equation}
where the integration is made over the dimensionless quantity $\eta={2^{2/3}x}/{\Lambda}$ and $\overline{\eta}>>1$.

This means that the focusing power $\overline{I_{F}}$ in (\ref{eq:initial_equation})
thus can be approximated reasonably well with the simple expression:

\begin{equation}
\overline{I_{F}}\sim \frac{1}{\Lambda}
\end{equation}

\subsubsection{Determination of the dead-zone position}

To find the propagation path of the focusing signal in an arbitrary
ionosphere, it is necessary in the context of wave optics to extract the dependence of the phase delay
 $\psi$ on the elevation angle $\alpha$. We use the fact that, at the focal points,
 the phase delay dependence on the elevation
angle is weakest, meaning that the derivative of the phase becomes zero \cite{Tinin_1991}.
Since we considered many possible ray paths, the focal points spread into a small area.
 Qualitatively, the size of such an area is a measure of the number
of rays arriving at the receiving point with the same phase, thereby adding to the total
amplitude of the received noise signal through constructive
interference of the rays.
 In terms of our model, this size
is inversely proportional to the cubic root of the second derivative
of the phase delay over the elevation angle (Eq.\ref{eq:lamda}) at the point where the
first derivative becomes zero (i.e. near the dead zone border), so that the power
of the focused signal is proportional to the size of this area \cite{Kravtsov_1968}.

It is obvious that in a complex ionosphere there can be several
focusing regions (modes). To find the positions (elevation angles) of the groundscatter
modes, we
analyzed all the possible propagation trajectories by calculating the
phase delay $\psi$ for each elevation angle $\alpha_i$.
After calculating the dependence of the phase delay on
the elevation angle $\psi(\alpha)$  the elevation angles $\alpha_{i}$ of the rays getting to the dead zones
were determined from the position where the phase delay reached its minimal
value:
\begin{equation}
\alpha_{i}:\left.\frac{d\psi(\alpha)}{d\alpha}\right|_{\alpha=\alpha_{i}}=0
\end{equation}
We should add that the local dependence $\psi(\alpha)$ was fitted to a parabola to make the evaluation of the second derivative
of  $\psi(\alpha)$ more easily to facilitate  the evaluation of $\Lambda$ (Eqn \ref{eq:lamda}).

We obtained a collection of angles $\alpha_{i}$ corresponding to different groundscatter modes.
In a regular, spherically layered, non-magnetized, parabolic ionosphere,
there is usually no more than one ground-scatter mode - there is only
one local minimum of the phase-elevation characteristic. In a more
complex but more realistic heterogeneous multi-layer ionosphere, 
there can be several groundscatter modes, with the  presence of
a magnetic field possibly adding extra possibilities.  Just with 
the propagation of ordinary rays, we found many instances of multiple  
ground-scatter modes, each with
its own phase, group delay and central elevation angle $\alpha$.

\begin{figure}
\includegraphics[scale=0.5]{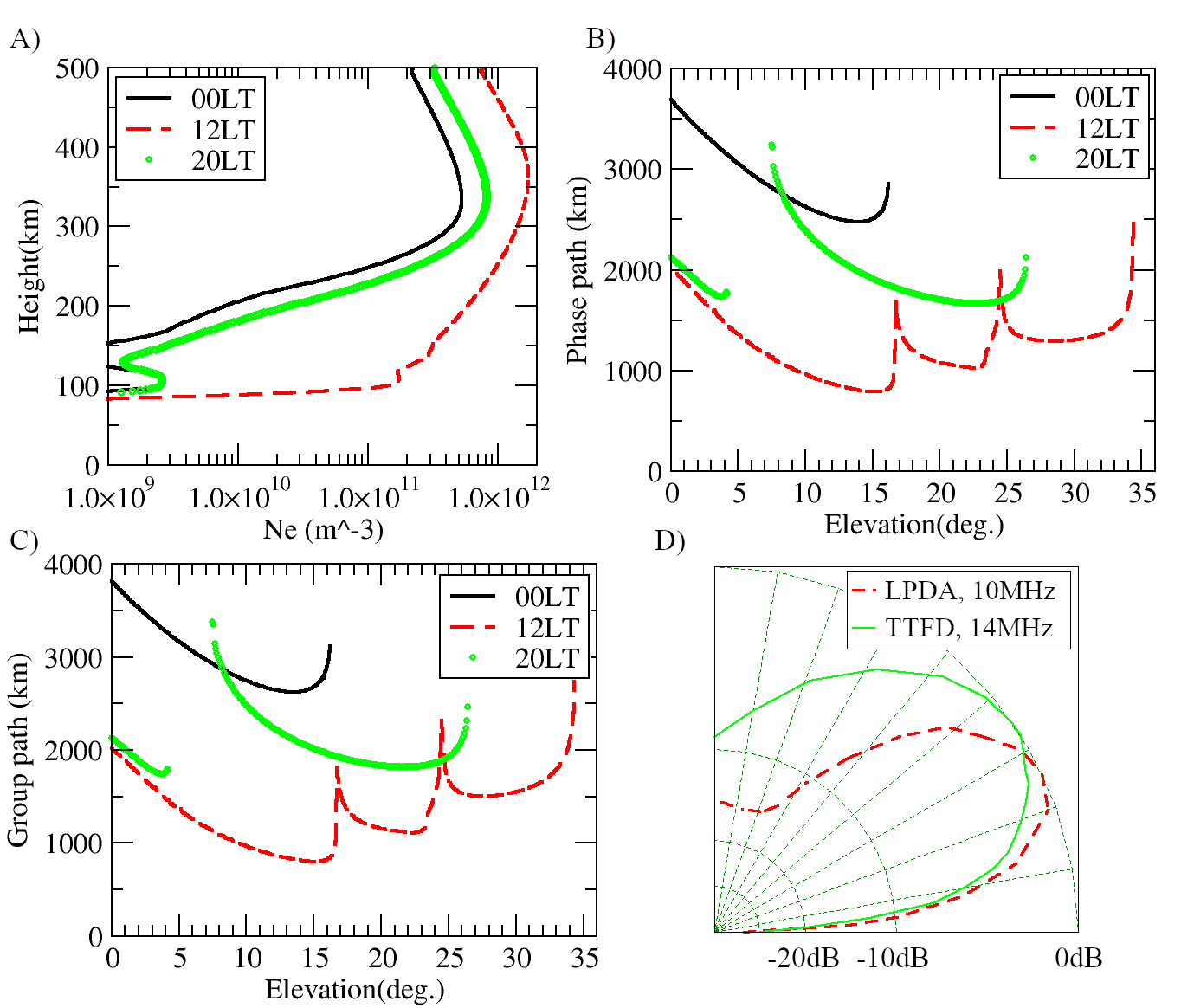}
\caption{ A-C)-Examples of electron density profile (A), phase path (B) and group
path (C), calculated using the IRI-2016 model for the EKB radar for the north-facing
direction on 01 June, 2013; D) - model radar antenna beam patterns in the elevation plane
used in our various simulations:
LPDA beam model at 10 MHz for the EKB, SAS and KER radars
\cite{Berngardt_2020b, Arnold_2003},
and TTFD beam model at 14 MHz for the BKS radar \cite{Sterne_2011}.
}
\label{fig:phaseDelayModel}
\label{fig:antenna_patterns}
\end{figure}

Examples of electron density profile at different local time(LT), as well as 
phase and group delay for single-mode
(black line) and multi-mode (red and green lines) signals, calculated
using the IRI-2016 model, are shown in Fig.\ref{fig:phaseDelayModel}A-C.
Modeling was carried out for the EKB radar (56.5N, 58.5E), for the
northern direction of radio wave propagation (azimuth 0, beam 2), for June
1, 2013. It can be seen from the figure that several modes correspond
to several well-defined (E and F) layers in the electron density.

When multi-mode noise trajectories were obtained, we chose the mode (trajectory)
that contributed the most to the noise signal intensity, after having taken  
signal absorption and antenna pattern attenuation into account.
Following this, the modeled noise level was chosen to be the intensity
of the mode with maximum intensity after taking
 all  factors into account, namely,  propagation, absorption, and antenna beam
pattern.

\subsection{Antenna pattern}

Another important parameter of the model is the signal attenuation
due to the antenna pattern $G_{A}$. The attenuation can reach several
tens of dB, and therefore it is very important to take it into
account at low and high elevation angles. In this paper, we have dealt with the
data from several radars: EKB from ISTP SB RAS, and  SAS, KER and BKS from SuperDARN.
Every one  of these radars has a different type of antenna and a different $G_{A}$.

Taking into account that the azimuthal width of the antenna beam pattern is defined by a phased array
geometry and size considerations (that much is similar for all the radars under consideration),
the integrated antenna multiplier $G_A$ has been taken to simply  be proportional to
the antenna pattern in the elevation plane, that is to say:
\begin{equation}
G_A(\alpha)=\int_{\Omega} G(\alpha,\Theta)d\Theta \approx G(\alpha)\Delta\Theta
\end{equation}
where $\Delta\Theta$ is the azimuthal beam width of the phased array antenna beam pattern and is the same for all the  radars considered here.  In addition,
$G(\alpha)$ is the antenna beam pattern of  a single antenna in the elevation plane.

Fig.\ref{fig:antenna_patterns}D shows the model antenna beam patterns
of the various radars at 10-14 MHz frequency that were used for our modeling.
It should be noted that the main difference between the various antenna beam patterns comes from the fact
that the EKB, SAS and KER radars use Log-Periodic Dipole Array antennas (LPDA) at the top of masts approximately
15 meters above the ground \cite{Arnold_2003,Berngardt_2020b}, while the BKS radar
 uses Twin Terminated Folded Dipole antennas (TTFD) at a lower height \cite{Sterne_2011}.
This significantly increases the amplitude of the vertical lobe on the
BKS radar and broadens its elevation antenna pattern as compared to the
EKB, SAS and KER radars.

\subsection{Noise absorption over the trajectory}

Another important parameter affecting the radio noise intensity is the
absorption of the signal over the propagation trajectory.
The absorption per unit length $L[dB]$ was calculated within the
framework of a classical model \cite{Zawdie_2017}
in which frequencies are considered to be high enough to justify neglecting  magnetic field effects:

\begin{equation}
L[dB]=-\frac{20}{log(10)}\frac{e^{2}}{2\epsilon_{0}m_{e}c}\frac{n_{e}\nu_{e}}{\nu_{e}^{2}+\left(2\pi f_{0}\right)^{2}}
\end{equation}

where, as before, $f_{0}$ is the operating frequency of the radar, $n_{e}$ is
the electron density, and  $m_{e}, c, \epsilon_{0}$ are the electron mass,
the speed of light in vacuum and the dielectric constant of  vacuum,
respectively while $\nu_e$ is the effective electron
collision frequency.  Since   $\nu_{e}$ is only effectual in the lower
part of the ionosphere we considered it to be equal to the  electron collision frequency with molecular nitrogen $\nu_{e,N_{2}}$.  Using
\cite{schunk_nagy_2000}, it was assumed  to be given by:

\begin{equation}
\nu_{e}\simeq \nu_{e,N_{2}}=2.33\cdot10^{-11}\cdot n_{N_{2}}(1-1.21\cdot10^{-4}T_{e})T_{e}
\end{equation}

where $n_{N_{2}}$ is the molecular nitrogen density and $T_{e}$ is
the electron temperature. The accuracy of this approximation has been discussed
in detail by \cite{Zawdie_2017}.

\begin{figure}
\includegraphics[scale=0.5]{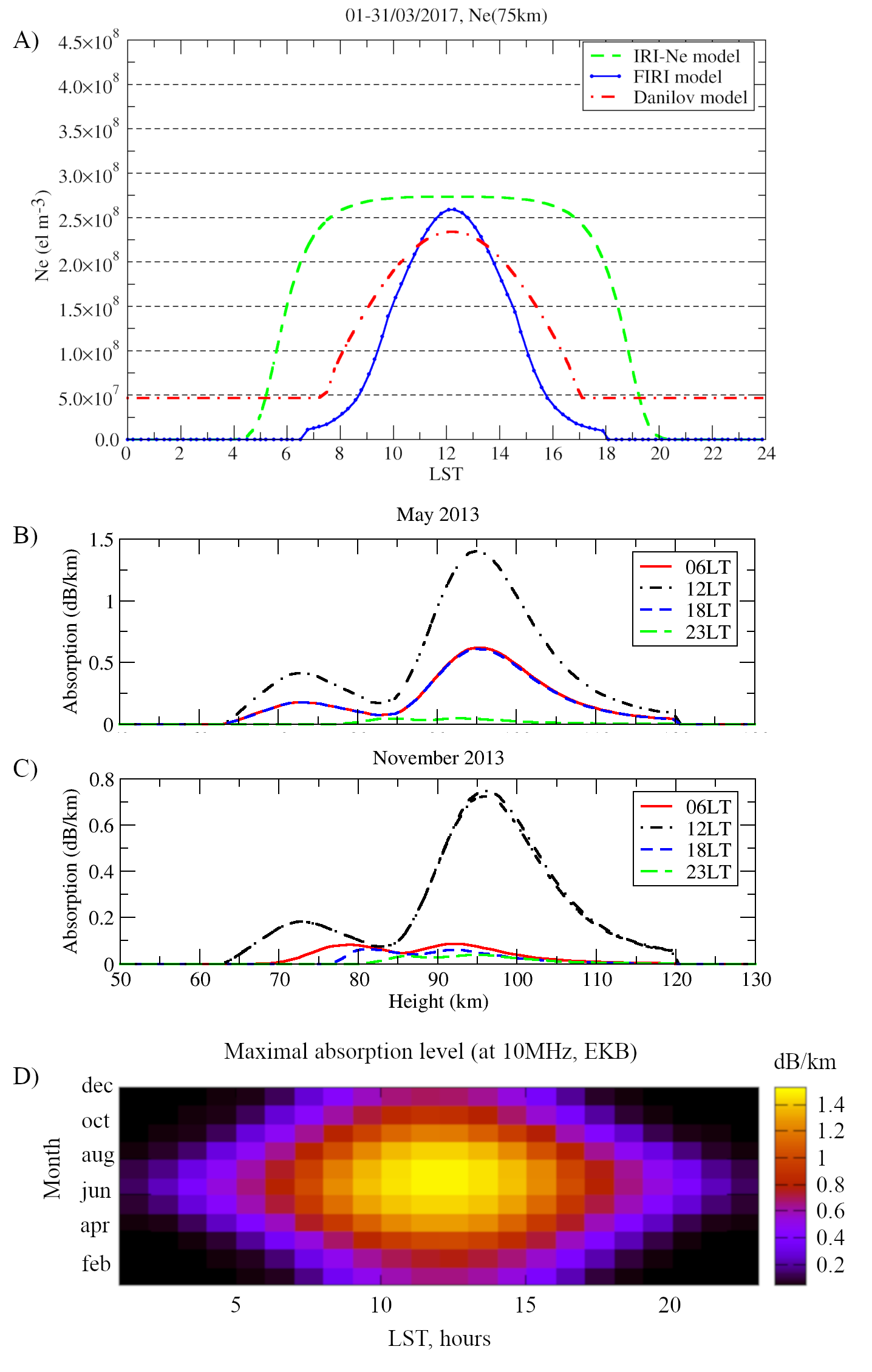}

\caption{ Panel A: Comparison between the monthly
average and median  electron densities at 75 km altitude extracted  
from the three different models listed in the legend.
Panels B to D: Modeling results for beam 2 of the EKB radar, 
beam 2 (0$^\circ$ azimuth) for the year 2013, based on the 
standard IRI-Ne option of the model IRI-2016.
Panel B:  modeled  absorption profile per 1 km propagation 
trajectory for May 2013.  Panel C: same as for panel B but for 
November 2013.  Panel  D:  seasonal-daily
variation of the maximum value of the absorption per 1 km of the
radio signal trajectory in the ionosphere.}

\label{fig:AbsorptionPerMeter}
\end{figure}

The total absorption of the radio wave $L_{int}[dB]$
was calculated as the integral of the unit absorption over the propagation
trajectory $S$, namely, \cite{Zawdie_2017}:
\begin{equation}
L_{int}[dB]=\int_{S} L[dB]ds
\end{equation}
In the simulations, the absorption was taken into account only for
the D- and E-layers, namely, in the 60-120 km altitude range, where 
significant contributions to the absorption were possible if $n_e$ was 
potentially large enough and where the collision frequency could compete with $f_0$.
For our model simulations, we describe the absorption factor $A_{t}$ in Eqn
(\ref{eq:initial_equation}) with
\begin{equation}
A_{t}=\exp\left(\frac{ln(10)}{10}L_{int}[dB]\right)\label{eq:A_t}
\end{equation}

\subsubsection{Weighting the factors that control the absorption}

To evaluate the absorption of the radio signal from the above expression, 
we used  well-known empirical models, namely,
the IRI-2016 model for a determination of  the electron
density $n_{e}$, and the NRLMSISE-00 model for a determination of 
the effective electron collision frequency based on the
density of molecular nitrogen $n_{N_{2}}$ and the electron temperature
$T_{e}$. For calculations between grid points, interpolation was performed
with two-dimensional local quadratic B-splines to obtain continuous
values required for trajectory calculations.
Then, in the plane of the propagation trajectory of the radio signal, the
absorption was tabulated over a grid which had 222 km range 
resolution and 5 km height resolution.  
In the simulations, interpolation based on local quadratic B-splines 
was once again used to account for 
a smooth spatial dependence.

One  source of uncertainty in the absorption calculation comes from 
accounting for the electron density in the  D-layer, which is no easy task, 
to say the least \cite{DANILOV2002,BILITZA2008}.
Even the standard Reference ionospheric model, IRI-2016, 
includes several different models for electron density at heights 65-110 km.
To demonstrate the complicated task at hand, 
Fig.\ref{fig:AbsorptionPerMeter}A
shows a comparison of the monthly average level of electron density
at 75 km height between three electron density models 
for the March 2017 (IRI, FIRI and Danilov's).
The IRI-2016 model comes with the standard IRI-Ne D-layer model.  
Other models of the D-layer come from \cite{DANILOV1995,MCKINNELL2006}.

Fig.\ref{fig:AbsorptionPerMeter}A shows that there are clearly important 
differences in the D region between the models.
In particular,  the nighttime  electron
density level could be significant (as demonstrated, for example, in \cite{Gomonov_2019}) 
and should be taken into account when
simulating the diurnal variation of the absorption. Taking this into
account leads us to conclude that the diurnal variation of the
absorption
inferred from the ionospheric models is associated not with the absolute value of
the electron density, but with its variation
over its minimum (nighttime) value. This means that the absolute electron density
of the  D-layer cannot be correctly determined through coherent scatter   radar absorption data
using noise level observations without  a very careful analysis of the  latter
and without an absolute calibration of these observations on a long term basis.  
A comparison between observations and numerical prediction is done in the next 
subsection.

Before a detailed comparison we want to  illustrate the effect of seasonal 
variations on absorption,     Figures \ref{fig:AbsorptionPerMeter}B and  
\ref{fig:AbsorptionPerMeter}C show
plots of the unit absorption per kilometer of the trajectory $L[dB]$ as 
a function of altitude that were calculated for
 May and November of 2013 to the North of the EKB radar (azimuth
0 degrees, radar beam 2), using the standard IRI-2016 model to describe the D-layer.
The radiowave propagation path used for the  calculation
corresponds to the trajectory of the
groundscatter signal propagation, i.e. to the trajectory from the dead zone boundary to the radar.
The absorption values are close, in the end, to the results obtained by \cite{Zawdie_2017}.

Figure \ref{fig:AbsorptionPerMeter}D also shows the full seasonal-daily variation
of the maximum value $L_{max}[dB]$ of absorption per km
of the trajectory in the ionosphere
\begin{equation}
L_{max}[dB]= Max\left(-L[dB]\right)_{h=[60..120km]}
\end{equation}
The calculation is based on the  standard IRI-2016 D-layer model for the EKB radar site. 
As can be seen, the maximum
absorption is observed in the summer under daytime conditions. This is the result 
of the ionization of the lower part
 of the ionosphere being greatest at that time of day and year.  It contrasts with 
winter and/or nighttime conditions when the lower ionosphere
 should have only weak ionization if any at all, according to  
the empirical models.

\section{Results and discussion}

\subsection{Comparison between simulations  and experimental observations}

\begin{figure}
\includegraphics[scale=0.4]{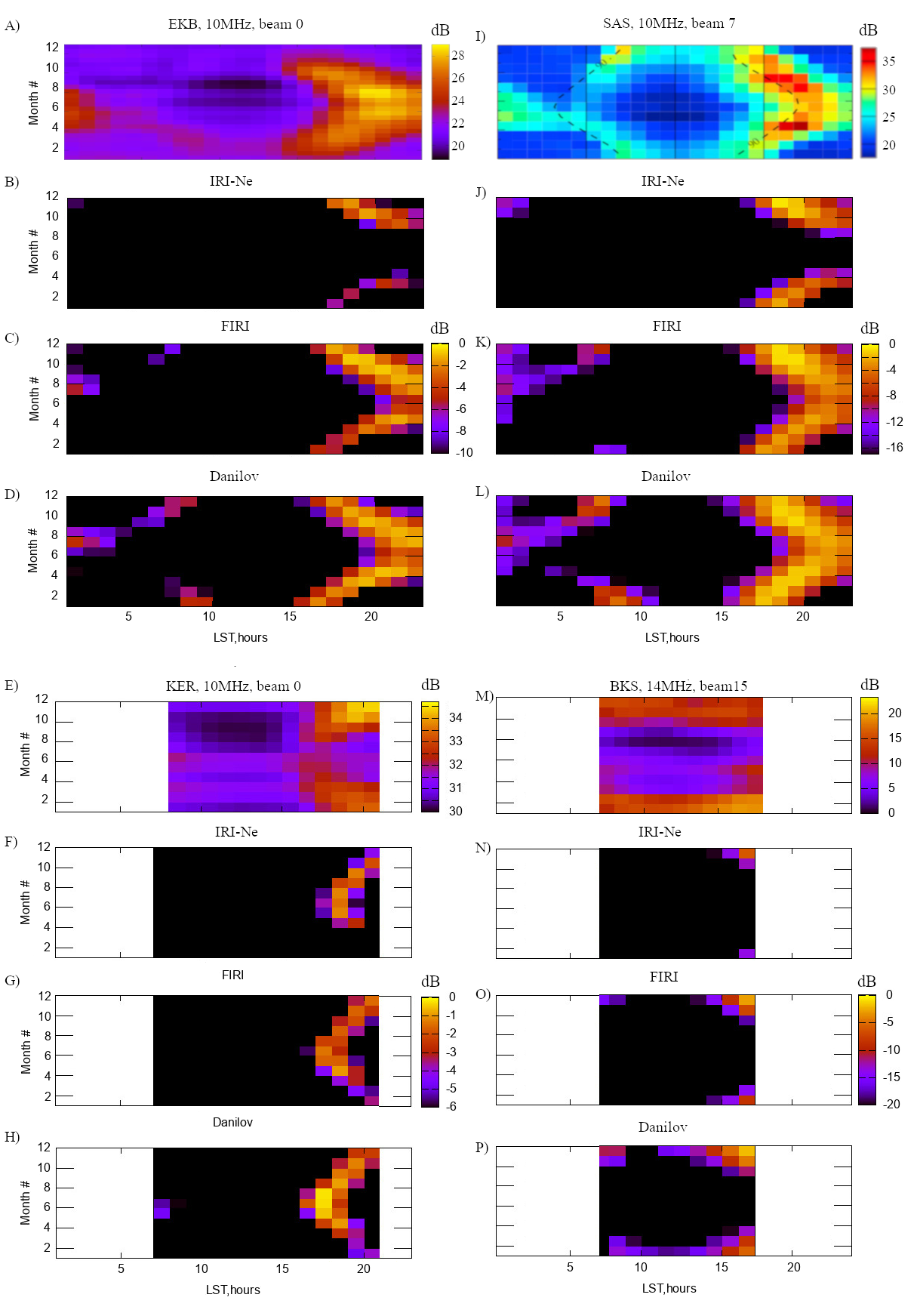}

\caption{Comparison between observations and simulations of noise level seasonal-daily dependence
for 2013 based on three different D-layer models, namely, 
IRI-2016, FIRI, and Danilov.
Panels A to D: EKB radar, 10MHz, at 0$^\circ$ azimuth with 
the top panel for observations and the next 3 panels for each model.
Panels E to H: same as panels A to D, but for the KER radar 
at 10 MHz for a 180$^\circ$ azimuth.
Panel I: SAS observations reported by \cite{Ponomarenko_2016}.
Panels J to L:  same as panels B to D, but for the SAS radar 
at 10 MHz for a $-23^\circ$  azimuth.
Panels M to P: same as panels A to D but for the  BKS radar 
at 14 MHz at $0^\circ$ azimuth.}

\label{fig:globalIRIcomparsionD}
\end{figure}

For the years 2013 and 2014, we have compared  the observed noise level for 
our four mid-latitude radars  with calculations based on our  noise level model
using three different D-layer models included into the international 
reference ionospheric model IRI-2016: (1)
the standard IRI-Ne \cite{Bilitza_1997} model, (2) the FIRI model \cite{MCKINNELL2006}
and (3) the Danilov model \cite{DANILOV1995}.  The comparisons are shown in
Fig.\ref{fig:globalIRIcomparsionD}.

A detailed comparison can be made with more confidence for the data 
associated with the  EKB and SAS radars, because the data from 
these radars came from  noise level observations
at a fixed frequency close to 10 MHz on a 24 hours per day basis.  
 From the results displayed in Figs.\ref{fig:globalIRIcomparsionD}A 
to \ref{fig:globalIRIcomparsionD}D, and
\ref{fig:globalIRIcomparsionD}I to \ref{fig:globalIRIcomparsionD}L 
for the EKB and SAS radars respectively, it can be seen that using 
standard IRI-Ne D-layer model
 leads to a significant overestimation of noise
absorption, while the FIRI and Danilov models lead to better qualitative
agreement between model calculations and observations in regions away from 
the regions of low noise prediction by the model (i.e., the uniformly dark 
central areas in the panels that describe the
computational results).

Data from the BKS and KER radars are harder to use for a comprehensive analysis
because they come from a variety of daytime and
nighttime sounding conditions. Nevertheless, the
daytime data that were retrieved (for 14 MHz at BKS, and for 10 MHz at KER) demonstrate
a good agreement with our model, particularly for the KER radar and  especially when using Danilov's model of the D-layer.

As one can see, the KER radar (southern hemisphere) seasonal-daily noise has a distribution that differs significantly
 from the northern hemisphere EKB, SAS and BKS radars.  This difference appears to come from the difference in hemispheres that are to be associated with different
  seasons and atmosphere/ionosphere
conditions. Indeed our model shows a good correspondence with the observations in
both the northern and southern hemispheres, implying that the model that we formulated based on Eqn. (\ref{eq:initial_equation}) has validity.

\subsection{Relationship between propagation parameters and noise variations}

\begin{figure}
\includegraphics[scale=0.45]{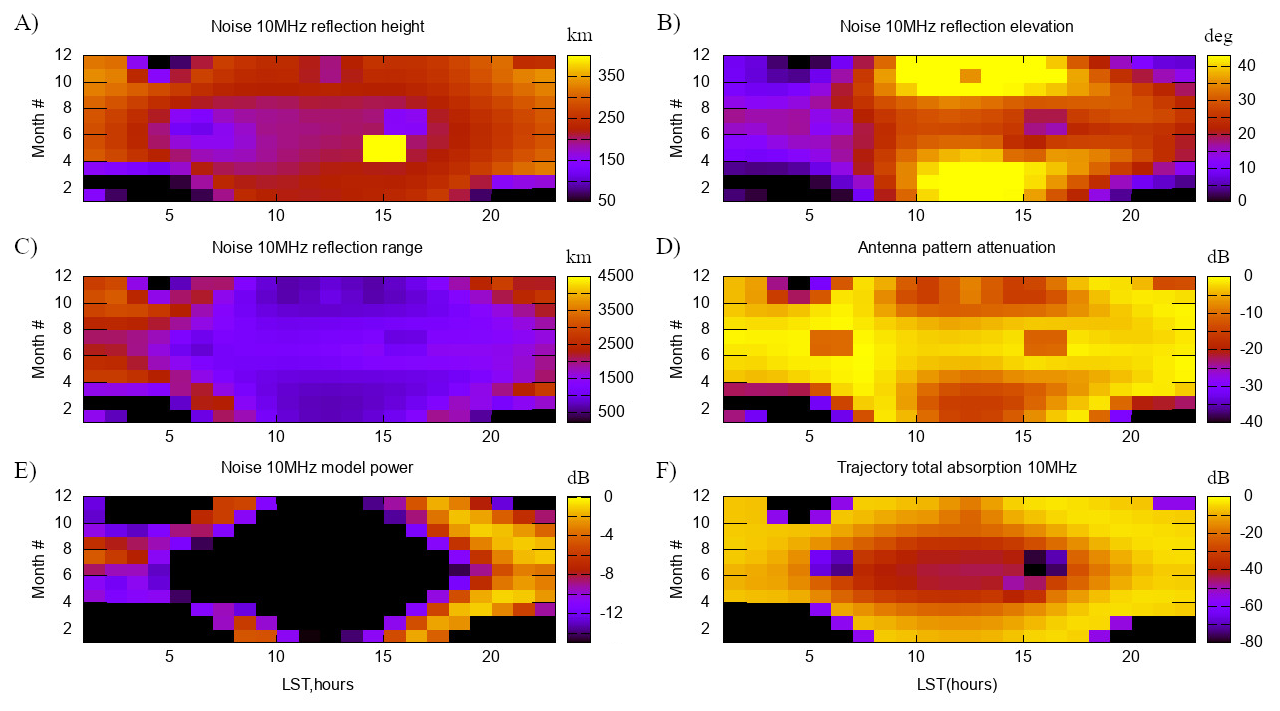}
\caption{Simulated seasonal-diurnal variation of  parameters associated with the propagation of radio waves for the EKB
radar in 2013 at 10 MHz. We used standard IRI-Ne model for D-layer simulation.}
\label{fig:ModelParametersAnalysis-10MHz}
\end{figure}

For a qualitative analysis, it is useful to demonstrate the relationship between the
propagation parameters and the seasonal-diurnal variation of the noise level.
Fig.\ref{fig:ModelParametersAnalysis-10MHz} shows the simulated results of seasonal-diurnal
variation of various propagation parameters at the EKB radar.
It can be seen from the figure that:

\begin{itemize}
\item the annular shape of the noise level (Fig.\ref{fig:ModelParametersAnalysis-10MHz}E)
is controlled mainly by the total absorption over the propagation trajectory
(Fig.\ref{fig:ModelParametersAnalysis-10MHz}F): in other words, the more 
absorption there is, the less noise is seen, with the two patterns matching very well;

\item the total absorption over the trajectory, in turn, is related to the
trajectory elevation angle (which changes the propagation
length of the signal in the absorbing layer) - the higher the elevation
angle, the shorter this length and the higher the noise level 
(Fig.\ref{fig:ModelParametersAnalysis-10MHz}B).

\item the total absorption is also related to the unit absorption in the regular
D- and E-layers - the stronger the absorption, the lower the noise
level(Fig.\ref{fig:ModelParametersAnalysis-10MHz}F versus \ref{fig:AbsorptionPerMeter}D). 
We should note that preliminary modeling has shown that the absorption below 85 km 
altitude is comparable to the absorption above 85 km;

\item the attenuation of the noise level in the winter daytime is associated
with antenna pattern attenuation at high elevation
angles(Fig.\ref{fig:ModelParametersAnalysis-10MHz}D versus Fig.\ref{fig:ModelParametersAnalysis-10MHz}B).
\end{itemize}

\subsection{Saturation effect and quiet day curve}

\begin{figure}
\includegraphics[scale=0.5]{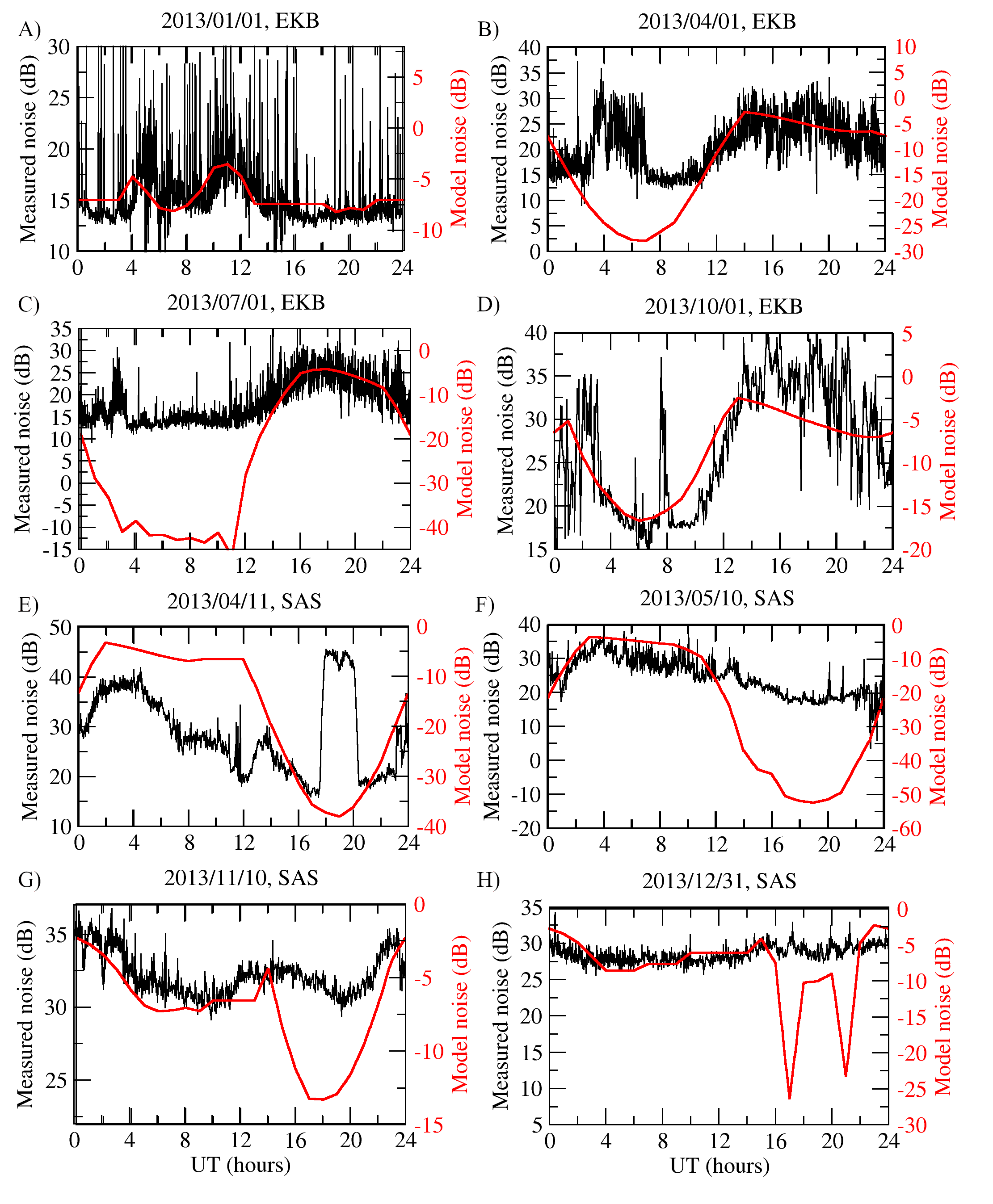}
\caption{Comparison of model (red line) and measured daily noise variations (black line) on radars
EKB (beam 2, A-D) and SAS (beam 7, E-H), at 10MHz}
\label{fig:EKB_SAS_example_cmp}
\end{figure}

As could be inferred from Fig.\ref{fig:ModelParametersAnalysis-10MHz} 
 the central region embedding summer daytime conditions seems to be  associated 
with noticeably weaker noise levels than
 expected from the model calculations.  The predictions for weak noise levels 
are based on strong absorption during sunlit conditions so that little noise 
from ground sources  should be expected to be found. We have
 carried out more detailed comparisons to assess this situation.  
For the purpose at hands, we focused on only two radars (EKB and SAS) 
because these were radars that used a fixed sounding frequency (10 MHz) 
on a continuous basis throughout the day. This
allows for a numerical comparison between the model and observations: 
Fig.\ref{fig:EKB_SAS_example_cmp}
shows some detailed examples of daily variations of the noise on the EKB
and SAS radars for different dates and their comparison with our proposed model
when we use the Danilov D-layer model for taking into account 
 absorption at lower altitudes, the IRI-2016 ionosphere
model for taking into account propagation effects, and NRLMSISE-00 neutral
atmosphere model for calculating the electron collision frequency.

As might have been expected from Fig.\ref{fig:ModelParametersAnalysis-10MHz}, 
the various panels in  Fig.\ref{fig:EKB_SAS_example_cmp} show a satisfactory 
agreement between the observations and
the model at high noise levels, but there is generally a strong disagreement 
when the model noise level is predicted to be  low.  We see two possibilities 
that could explain this situation.
One is that there may well be a saturation effect associated
with analog and digital sensitivity levels of the radars.  Alternatively, 
or even in association with instrumental noise, cosmic noise could be a factor, 
as it can sometimes  be as high as 20 dB
below anthropogenic noise, as shown in \cite{ITU_R}(Figs.2,39).  
Cosmic noise comes from much higher elevation angles and
could make a significant contribution when anthropogenic noise 
is not too strong because of absorption.  We have not queried this question 
any further as the focus of the present
paper is anthropogenic sources from the ground.  Our model of these contributions 
produces the bright arc patterns in Fig.\ref{fig:globalIRIcomparsionD}.  
The important message is that these arcs fit
well with the notion of noise from anthropogenic sources.

\begin{figure}
\includegraphics[scale=0.8]{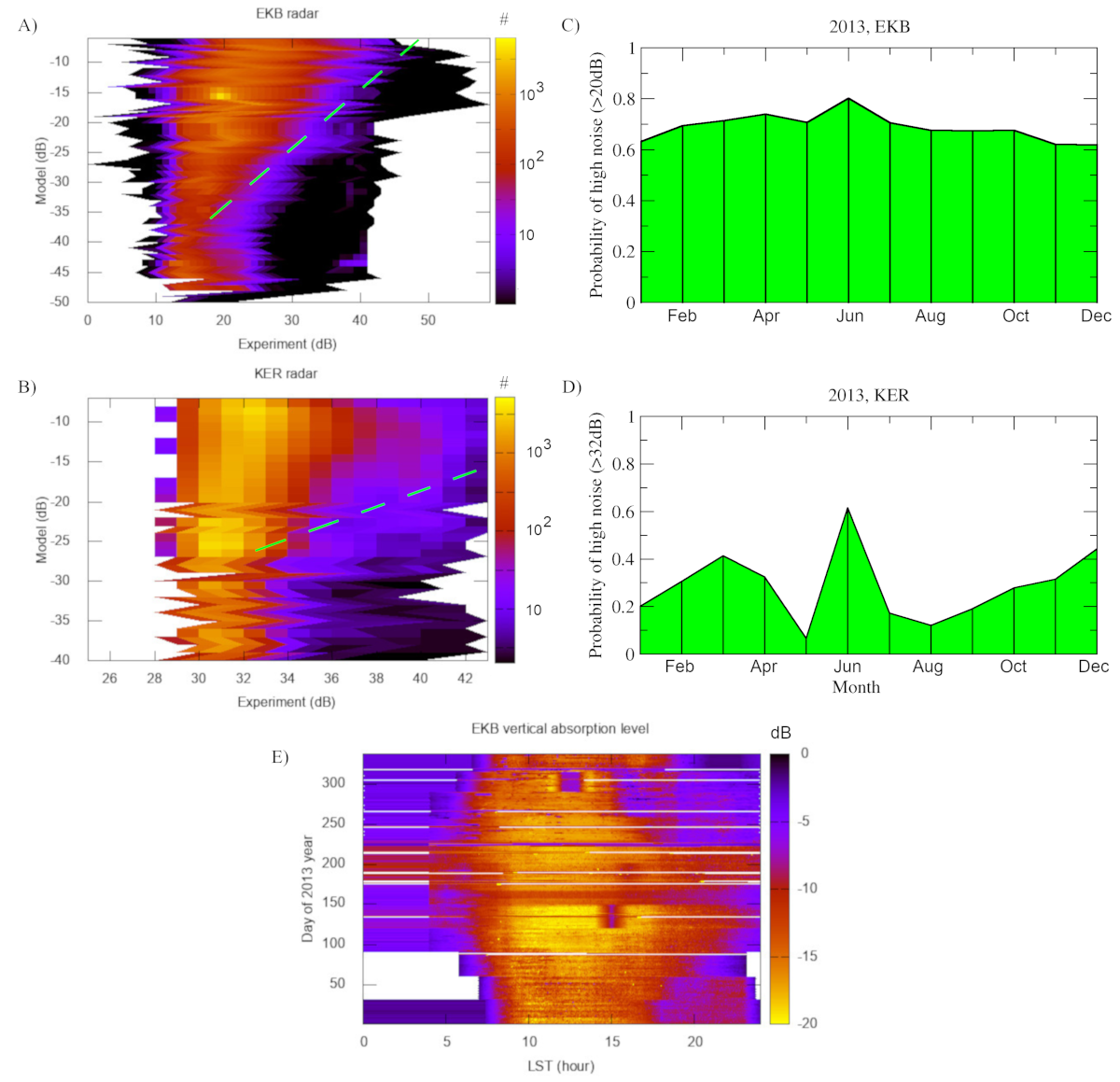}
\caption{Comparison between model and experimental data for the EKB (A)  and KER (B) radars
and monthly statistics of radar data available for adequate numerical
analysis within the framework of the proposed model: for EKB (C) and
KER (D) radars.
Dashed line in (A,B) corresponds to linear dependence, 
the color indicates the corresponding number of occurences of combinations of 
experiment vs. model values.
E) Vertical absorption at 10MHz calculated from experimental EKB noise data using the suggested model.
}
\label{fig:EKB_sesnsittivity_effects}
\end{figure}

To demonstrate that saturation effects are very real, we introduce Fig.\ref{fig:EKB_sesnsittivity_effects}.
Fig.\ref{fig:EKB_sesnsittivity_effects}A and B show the results
of a statistical comparison between the model and the observations using
the EKB and KER radar data in the year 2013. 
The color shows the number of cases when a particular 
combination of experimental and model noise values was obtained.

Consider  Fig. \ref{fig:EKB_sesnsittivity_effects}A first: clearly, 
when the observed noise level is less than 20 dB, there is 
no connection between the predicted
noise level and the observed one. However, when the observed 
noise level is greater than 20 dB there is a linear relation  
between prediction and observations.
From this it should be clear that the  measured EKB noise levels
are saturated  below 20 dB, thereby having nothing to do 
with ionospheric absorption.  By contrast, above 20 dB, 
the effects of ionospheric absorption are seen.
The clear implication is that the model cannot be used below 20 dB 
in the EKB case because of noise saturation effects affecting the observations.

Fig.\ref{fig:EKB_sesnsittivity_effects}B shows that a similar effect 
takes place for the KER radar. In that case, however,
the saturation level is higher (32 or 33 dB),
and the physical information that can be extracted from  noise variations 
is for noise levels that are weaker by about 8 to 10 dB.

 Fig.\ref{fig:EKB_sesnsittivity_effects}C and D display similar 
information in a somewhat different way through monthly statistics.  For those panels,
 we kept track of the radar data that could be suitably compared with a
numerical interpretation of the observations by the proposed model 
(as per the numbers inferred from  panels A and C, meaning a noise
level above 20 dB for the EKB radar and above 32 or 33 dB for the KER radar). 
This being the case,   Fig.\ref{fig:EKB_sesnsittivity_effects}C shows that 
the probability that the noise level at EKB be  20 dB above the numbers 
expected from our simulations   is always above 60\% at all times of 
the year, with a somewhat higher value (closer to 80\%) in June.  
The numbers are different for KER, as seen from 
Fig.\ref{fig:EKB_sesnsittivity_effects}D. We note that the statistics 
are too weak to give any significance to the June peak in that case 
(a change from 32 dB to 33 dB would put the peak well inside a region 
of minimum probability).

From our analysis of Fig.\ref{fig:EKB_sesnsittivity_effects}A to D
we may conclude  that about a half
of the daily noise data at EKB  is suitable for the numerical interpretation. 
These periods where the simulations work reasonably well
 are usually associated with a weak D-layer ionization level. They come
 mainly from nighttime and autumn-winter months. This much is clear after we take 
another look at Fig.\ref{fig:globalIRIcomparsionD}.
The smaller size of the dataset in KER is  caused by its smaller dynamical range
(Fig.\ref{fig:EKB_sesnsittivity_effects}B), itself most likely caused by a lower level of
distributed surface anthropogenic sources around the radar (Fig.\ref{fig:fig1}C)
, which is isolated in the Southern Indian Ocean.

With the above reservations concerning extra noise sources in mind, 
the proposed model can be used to build a quiet day curve (the results
of which are presented through the red traces in
Fig. \ref{fig:EKB_SAS_example_cmp}).  However, for the model to be 
put to good use, we need to take
into account the saturation effect of the radar sensors
at low noise levels.  The saturation level clearly  differs for 
each specific radar (Fig.\ref{fig:EKB_sesnsittivity_effects}A,B)
, and its mechanism and dynamics should be studied in the future.
As shown above,  this model depends on a number of factors, namely, (1) the
trajectory of propagation of the noise signals (or ground-scatter signals),
(2) the antenna pattern in the elevation plane of propagaion, as well as (3) the mode
of the received signal as it is affected by propagation in a three-dimensionally
inhomogeneous ionosphere. The model used for the `quiet' ionospheric D-layer 
is also significant in the construction of the quiet
day curve. The comparison of observations with model simulations 
(Fig.\ref{fig:globalIRIcomparsionD}) shows that the use of the traditional 
IRI-2016 model leads to significant overestimation of the noise level.
The Danilov model of the D-layer leads to  better fits to observed noise 
levels than the other models, at least  for  the EKB radar 
(Fig.\ref{fig:globalIRIcomparsionD}).

\subsection{Implications for the modeling of ionospheric absorption}

Based on the use of Eqn.(\ref{eq:initial_equation}) for the construction of our model, 
one can exclude  all propagation effects
(antenna pattern attenuation $G_{A}$,
radio wave attenuation with distance $\overline{A_R}$,
and focusing effects $\overline{I_{F}}$ )
on the  noise level variations $\overline{I_{N,rcv}}$ as well as exclude attenuation 
variations in the elevation angles through  propagation
for a trajectory through an angle $\alpha$ in the D-layer
(valid for a thin non-refracting absorption layer).   In other words the intensity 
is essentially 'fixed' through the relation

\begin{equation}
I_{N,fxd}[dB]=10 \cdot \log_{10} \left(
\frac {\overline{I_{N,rcv}}} { I_{0,N}\overline{I_{F}}G_{A}\overline{A_R}}
\right)
\end{equation}

 while the vertical absorption becomes 

\begin{equation}
A_{t,vert}[dB]=(I_{N,fxd}[dB]-Max(I_{N,fxd}[dB])) \cdot \sin(\alpha)
\end{equation}

where the maximum value is calculated over the year to calibrate absorption when it reaches a minimum.
As a result,
one can obtain an estimate for the equivalent vertical absorption $A_{t,vert}$ near the radar.
Fig.\ref{fig:EKB_sesnsittivity_effects}E shows the seasonal-daily dynamics of the vertical absorption
$A_{t,vert}$
near EKB  based on this approach to the analysis of the noise observations at 10 MHz.
As one can see, the vertical absorption
goes through a maximum at summer daytime.  This corresponds well, qualitatively, to the D-layer absorption
expected from the IRI-2016 and NRLMSISE-00 models, as shown in Fig.\ref{fig:AbsorptionPerMeter}D.

\section{Conclusion}

In this paper, we have developed a numerical  model of
the seasonal-diurnal dynamics of the radio noise at polar-oriented HF-radars.  The  model has focused on
 four high-frequency mid-latitude
coherent scatter radars - EKB from ISTP SB RAS, and
SAS, BKS and KER from SuperDARN.  The numerical results  have been compared with observations.

The results from our simulations give credence to the model proposed earlier in \cite{Berngardt_2018,Berngardt_2019}
for interpreting the radio noise level in terms of the spatial
focusing of ground sources of anthropogenic origin in the vicinity of the `dead zone' propagation boundary.
Using the IRI-2016 and NRLMSISE-00 models, it has been  shown here that seasonal-diurnal
variations of the noise level can be explained within the framework
of this model through a combination of propagation, absorption and antenna
pattern factors.

We have demonstrated that the seasonal-diurnal patterns  observed in radio noise 
(with a maximum in the region of the solar terminator) is
associated with variations in the total absorption of the signal over
the propagation trajectory to the border of the dead zone. We have also shown
that an observed decrease in the noise intensity in the winter noon area in the EKB
and SAS radars (not detected with the BKS radar) can be explained
by an additional attenuation of the
signal by the antenna beam pattern at high elevation angles.

The agreement with observations is particularly good near the terminator 
where the noise level is greatest.  
This has allowed to us compare the absorption computed from different 
D region models and compare the resulting noise levels with observations.
We found that the absorption levels, when integrated along the ray paths, 
were too high when using 
the standard IRI-2016 D-layer model.  However, 
the agreement with observations based on the use of the FIRI \cite{MCKINNELL2006}
and Danilov \cite{DANILOV1995} D-layer models was much better.

The above notwithstanding, in spite of a good qualitative agreement of the seasonal-diurnal
model variations with observations, quantitative simulation
results demonstrate that the model predicts far less noise than observed away from 
the terminator during sunlit conditions.
There are only two possible explanations: either the model produces too much absorption, 
or there is another source of noise that
does not involve ground-based anthropogenic sources. In the former case, one would have 
to face the fact that the model produces realistic amounts of absorption at other times,
 particularly near the terminators.    In the latter case there are two possibilities, 
both of which must be present but were not considered for our model based on ground sources: 
cosmic background noise, which comes from above and would
favor larger elevation angles, and instrumental noise, i.e., saturated noise counts.  
The ability to measure absorption variations from radar noise measurements depends on 
the surrounding anthropogenic noise level - the higher the noise level, the stronger absorption 
events the radar can study. For radars with low anthropogenic noise level, such as KER, 
it may be difficult to study strong vertical absorption events from the noise data.

The main message to carry from the present study is that 1) the mechanism proposed in 
\cite{Berngardt_2018,Berngardt_2019} does very well near the terminator where the noise 
level is at its strongest Also, 2) 
there are additional noise sources to worry about, be they instrumental or of cosmic origin. 
Finally 3) the present model should provide a useful tool for the assessment of absorption 
on a day-to-day basis.
This means that we could use the present numerical model or its successors to extract 
important information from the dynamics of noise measured by pole-oriented coherent
scatter radars in the 8 to 20 MHz range.

It should be kept in mind that the results presented here can only be 
considered as preliminary, as they use
reference ionospheric model IRI-2016 to calculate the propagation trajectory. 
For more correct calculations
of the propagation path one really ought  to take into account  actual ionospheric refraction,
estimated either from  radar data
(using, for example, ground-scatter signals in combination with  measured elevation angles) 
or from ionospheric electron
density measurements by ionosondes. This approach will be considered in  future studies.

\acknowledgments
The data of EKB radar were obtained using the equipment of Center for Common Use
'Angara' http://ckp-rf.ru/ckp/3056/ and available at http://sdrus.iszf.irk.ru /ekb /page\_example /simple.
The work of EKB radar was financially supported by the Ministry of Science and 
Higher Education of Russian Federation.
Work of OB and AM was supported by joint RFBR-CNRS grant \# 21-55-15012 NCNI\_a.  
JPSM receives support from the Canadian NSERC.
The authors acknowledge the use of SuperDARN data.  
SuperDARN is a collection of radars funded by national scientific funding agencies 
of Australia, Canada, China, France, Italy, Japan, Norway, South Africa, United Kingdom 
and the United States of America. 
We thank all participants in the worldwide SuperDARN collaboration for the
distribution of SuperDARN data via
http://vt.superdarn.org/ tiki-index.php?page=Data+Access.
The Saskatoon SuperDARN radar
operations are specifically funded by Canadian Space Agency contracts and by a
Major Science Initiative from Canada Foundation for Innovation.
The SuperDARN Kerguelen radar is operated by IRAP 
(CNRS, Toulouse University and CNES) and is funded by 
IPEV and CNRS through INSU and PNST programs.
The SuperDARN Blackstone radar is operated by Virginia Tech 
with funding from the National Science Foundation under award AGS-1935110.


%
%

\bibliography{refs}

%
%
%
%
%

\end{document}